\pdfoutput=1
\documentclass[letterpaper,titlepage,11pt]{article}

\usepackage{jheppub} 
\usepackage[T1]{fontenc}
\usepackage[utf8]{inputenc}
\usepackage[turkish, english]{babel} 

\usepackage[active]{srcltx}
\usepackage{textcomp}
\usepackage{amsmath}
\usepackage{amssymb}
\usepackage{esint}
\usepackage{multirow}
\usepackage{empheq}
\usepackage{textcomp}
\usepackage{xcolor}

\usepackage{float}
\usepackage{booktabs}

\def\cA{\mathcal{A}}

\def\cF{\mathcal{F}}

\def\cL{\mathcal{L}}
\def\cM{\mathcal{M}}
\def\cN{\mathcal{N}}

\def\cW{\mathcal{W}}
\def\cX{\mathcal{X}}
\def\cY{\mathcal{Y}}

\usepackage{enumerate}
\usepackage{xcolor}
\def\be{\begin{eqnarray}}
\def\ee{\end{eqnarray}}
\def\beann{\begin{eqnarray*}}
\def\eeann{\end{eqnarray*}}
\def\beq{\begin{equation}}
\def\eeq{\end{equation}}
\def\ba{\begin{array}}
\def\ea{\end{array}}
\def\ben{\begin{enumerate}}
\def\een{\end{enumerate}}
\def\bea{\begin{eqnarray}}
\def\eea{\end{eqnarray}}

\providecommand{\Lt}{{\tt L}}
\renewcommand{\Lt}{{\tt L}}
\providecommand{\Wt}{{\tt W}}
\renewcommand{\Wt}{{\tt W}}

\def\cA{{\cal A}}

\def\cF{{\cal F}}

\def\cL{{\cal L}}
\def\cM{{\cal M}}
\def\cN{{\cal N}}

\def\cW{{\cal W}}
\def\cX{{\cal X}}
\def\cY{{\cal Y}}

\def\be{\begin{equation}}
\def\ee{\end{equation}}
\def\bea{\begin{eqnarray}}
\def\eea{\end{eqnarray}}
\def\ba{\begin{array}}
\def\ea{\end{array}}

\usepackage{amsfonts}

\usepackage{tikz}

\usepackage{bbm}

\DeclareMathOperator{\extdm}{d}
\newcommand{\extd}{\extdm \!}

\title{\center{\bf{On the Explicit Asymptotic  Symmetry Breaking of $sl(3,\mathbb{R})$ Jackiw$-$Teitelboim Gravity}}}

\subheader{ITU Preprint,\,\\ \today}
\newcommand{\itu}{\dagger}

\author[\itu]{H. T. \"Ozer}
\emailAdd{ozert@itu.edu.tr}
\author[\itu]{,\,\,\,Ayt\"ul Filiz}
\emailAdd{aytulfiliz@itu.edu.tr}

\affiliation[\itu]{Istanbul Technical University,\,Faculty of Science and Letters,
\,Physics Department,\\34469 Maslak,\,Istanbul,Turkey.}
\abstract{
This study investigates the asymptotic symmetry algebras (ASA) of Jackiw--Teitelboim (JT) gravity
 within the framework of \(\mathfrak{sl}(3,\mathbb{R})\) symmetry. By explicitly constructing this algebra, 
 we explore how the presence of the dilaton field influences the structure of asymptotic symmetries
  and symmetry breaking mechanisms at the AdS$_2$ boundary. For the \(\mathfrak{sl}(3,\mathbb{R})\) model, the 
  dilaton field preserves a subset of the complete $W_3$--symmetry, restricting the algebra to 
  \(\mathfrak{sl}(3,\mathbb{R})\). These results provide deeper insights into the role of dilaton dynamics 
  in holographic dualities, with implications for the thermodynamics and geometry of AdS$_2$. 
  The findings pave the way for systematically exploring extended gauge symmetries in 
  two--dimensional gravity and their relevance to higher--rank Lie algebras.
}


\begin{document}

\maketitle
\flushbottom
\section{Introduction}
\label{sec:intro}
\vspace{0.5cm}

The holographic principle~\cite{Maldacena:1997re} is known as a fundamental concept suggesting that a 
$d$-dimensional theory of gravity can be equivalent to a $(d{-}1)$-dimensional boundary theory. In this 
context, the study of $\mathrm{AdS}_2$ holography in two dimensions began in the late 1990s
~\cite{Strominger:1998yg,Cadoni:1998sg,Hotta:1998iq,Navarro-Salas:1999zer}, focusing on the asymptotic 
symmetries of Jackiw–Teitelboim (JT) gravity~\cite{Teitelboim:1983ux,Jackiw:1984je} and dilaton models 
obtained via dimensional reduction. In recent years, interest in this subject has been revived by the 
proposal that JT gravity is holographically dual to the Sachdev–Ye–Kitaev (SYK) model
~\cite{Maldacena:2016hyu,Jensen:2016pah,Sachdev:1992fk,Kitaev}; within this framework, the Schwarzian 
boundary action emerges as an intermediate description~\cite{Mertens:2018fds}.
\vspace{0.5cm}

Conformal field theory (CFT) has become a central framework in modern theoretical physics, with applications 
ranging from string theory to condensed matter systems. Its importance largely arises from the strong symmetry 
constraints it imposes, which significantly restrict the form of physical theories. In two dimensions, conformal 
symmetry is especially powerful, giving rise to an infinite-dimensional structure known as the Virasoro algebra
~\cite{Virasoro:1969zu}. Over the years, considerable attention has been devoted to extending this symmetry to 
include more general algebras, such as the Kac–Moody algebra~\cite{Kac_1968,Moody:1968zz} and the $W$-algebra
~\cite{Zamolodchikov:1985wn}. At the same time, many studies have focused on a deeper understanding of the Virasoro
algebra itself~\cite{DiFrancesco:1997nk,Blumenhagen:2009zz,Prochazka:2014gqa,Ozer:2015dha}. These symmetry 
structures also play a central role in the analysis of two-dimensional dilaton gravity, where the emergence 
of the Virasoro algebra at the boundary provides a natural bridge between CFT techniques and gravitational dynamics.
\vspace{0.5cm}

Two-dimensional pure gravity can be formulated as a topological theory with no propagating degrees of freedom. 
In the AdS$_2$ case, the system exhibits an \( SL(2,\mathbb{R}) \) isometry group, which also serves as its 
asymptotic symmetry group, resulting in a vanishing central charge~\cite{Almheiri:2014cka}. However, introducing 
a dilaton field renders the system dynamical, leading to models such as Jackiw–Teitelboim (JT) gravity
~\cite{Teitelboim:1983ux,Jackiw:1984je}, where the dilaton couples to the metric and introduces additional 
degrees of freedom at the boundary~\cite{Maldacena:2016upp,Jensen:2016pah}. In this framework, the asymptotic 
symmetry group extends beyond \( SL(2,\mathbb{R}) \) to the full infinite-dimensional Virasoro algebra
~\cite{Cadoni:1999ja}, allowing for a nonzero central charge. JT gravity has gained prominence as an 
effective model for probing aspects of quantum gravity, particularly in connection with the SYK model, where the 
Schwarzian action governs boundary dynamics ~\cite{Maldacena:2016upp,Engelsoy:2016xyb,Almheiri:2014cka}. 
These developments have elevated two-dimensional  dilaton gravity to a central role in exploring holography, 
black hole thermodynamics, and quantum information
~\cite{Cvetic:2016eiv,Sachdev:2019bjn,Sachdev:2010um}.
\vspace{0.5cm}

The rich structure of JT gravity lies in its boundary symmetries and mathematical formulation. The model is deeply 
connected to conformal symmetry via the Virasoro algebra and Schwarzian derivatives~\cite{Iliesiu,Mertens,OvisenkoSchwarzian}, 
and has been extended through relaxed boundary conditions and higher-spin generalizations~\cite{Gross:2016kjj,Witten:2016iux,Yoon}. 
Boundary dynamics can be expressed in terms of coadjoint orbits, providing insight into symmetry breaking mechanisms and their physical consequences~\cite{Davison:2016ngz,NarayanYoon,Fu:2016vas,Grumiller:2017qao,Alkalaev:2013fsa,Grumiller:2013swa,Grumiller,Kine,Fukuyama}. 
While the Schwarzian formalism remains central to the SYK connection, the analysis of asymptotic symmetries can proceed 
independently, for instance via the BF formulation of JT gravity~\cite{Grumiller:2002nm,Banks:1990mk}. In this approach, 
the dilaton acts as a Lagrange multiplier imposing curvature constraints, and the resulting asymptotic symmetry algebra 
can be enhanced—e.g., to Virasoro–Kac–Moody—through internal symmetry structures~\cite{Ikeda:1993fh,Jackiw:1982hg}. This 
flexibility highlights the utility of JT gravity in addressing low-dimensional gravitational dynamics from multiple perspectives, 
with or without reliance on Schwarzian dynamics. 
\vspace{0.5cm}

In the context of two-dimensional JT gravity, the Schwarzian formalism is commonly employed to describe boundary dynamics 
and establish a connection with the SYK model~\cite{Grumiller:2002nm, Ikeda:1993fh, Jackiw:1982hg}. However, when the primary 
goal is to analyze asymptotic symmetries rather than boundary behavior, this formalism is not essential~\cite{Grumiller:2002nm, Jackiw:1982hg}. 
Analogous to the treatment of three-dimensional gravity via the Chern–Simons formulation and Brown–Henneaux boundary conditions, 
JT gravity can be formulated as a BF theory in which the dilaton acts as a Lagrange multiplier enforcing curvature constraints
~\cite{Grumiller:2002nm, Banks:1990mk}. This approach allows for the emergence of an extended asymptotic symmetry algebra—typically 
Virasoro or even Virasoro–Kac–Moody—through the inclusion of internal symmetry structures and the application of appropriate boundary 
conditions~\cite{Ikeda:1993fh, Jackiw:1982hg}. Coadjoint orbit methods~\cite{Ikeda:1993fh, Jackiw:1982hg} offer a systematic 
framework for exploring these symmetries, particularly in cases where symmetry breaking at the boundary plays a crucial role. 
Thus, this formulation highlights the flexibility of JT gravity in addressing holographic and asymptotic structures beyond the Schwarzian regime.
\vspace{0.5cm}

Moreover, a natural classification of boundary conditions in AdS$_2$ gravity distinguishes between the most general (affine) and the 
conformal types, each corresponding to distinct asymptotic symmetry algebras and dual dynamics. The most general conditions are naturally 
formulated within the Poisson sigma model framework and allow for both the dilaton and the boundary metric to fluctuate at leading order. 
This leads to an infinite-dimensional asymptotic symmetry algebra given by a centerless affine $\mathfrak{sl}(2,\mathbb{R})$ current algebra. 
While the Schwarzian action can be obtained as a particular projection within this setup, the full theory goes well beyond it. In contrast, 
the conformal boundary conditions fix the boundary geometry and induce a Schwarzian effective action at low energies, capturing the dynamics 
of the pseudo-Goldstone modes associated with the breaking of reparametrization symmetry. These conformal conditions give rise to an off-shell
Virasoro symmetry, spontaneously broken to $\mathrm{SL}(2,\mathbb{R})$ on-shell, and are known to provide the gravitational dual of the IR 
sector of the SYK model. Together, these two boundary conditions represent complementary regimes in the study of AdS$_2$ holography, 
highlighting the versatility and richness of two-dimensional dilaton gravity beyond the Jackiw–Teitelboim model.
\vspace{0.5cm}

In contrast to the well-studied $sl(2,\mathbb{R})$ formulation of JT gravity, where the boundary symmetry is governed by the linear Virasoro algebra, 
the $sl(3,\mathbb{R})$ extension provides a natural setting for exploring richer, nonlinear structures such as the $\mathcal{W}_3$ algebra. 
This choice is not arbitrary but arises from the desire to investigate higher-spin generalizations of dilaton gravity and their implications 
for boundary dynamics. In particular, the inclusion of spin-3 fields leads to a deformation of the asymptotic symmetry algebra, where the 
dilaton fields not only extend the symmetry but also induce controlled symmetry breaking. This framework offers new insights into the behavior 
of two-dimensional gravity under generalized boundary conditions and contributes to ongoing efforts to connect higher-spin JT gravity with 
holographic duals such as SYK-like models.  Moreover, since $\mathcal{W}_3$ symmetries naturally arise in large-$N$ limits of matrix models 
and certain generalizations of the SYK model, our framework may provide a useful holographic dual for such deformed CFT$_1$ theories. 
We comment on the implications for boundary correlators and charge algebras in Section~\ref{sec:conclusion}.
\vspace{0.5cm}

While the $\mathfrak{sl}(2,\mathbb{R})$ JT model features a reduction from the infinite--dimensional Virasoro algebra to $\mathrm{SL}(2,\mathbb{R})$ 
due to boundary dilaton fluctuations, our extended model exhibits a similar mechanism for the breaking of the $\mathcal{W}_3$-- algebra. The dilaton 
and higher-spin fields serve as dynamical stabilizers that constrain the asymptotic gauge transformations, effectively reducing the full 
$\mathcal{W}_3$-- symmetry to its $\mathrm{SL}(3,\mathbb{R})$ subalgebra. This provides not only a consistent truncation but also a pathway 
to study symmetry breaking in higher-spin holography within a BF-theoretic framework.  Here and throughout this work, the $\mathrm{SL}(3,\mathbb{R})$  
subalgebra expresses the extended form of the affine $\mathfrak{sl}(3,\mathbb{R})_k$ or conformal $\mathfrak{sl}(3,\mathbb{R})$   $\mathcal{W}_3$--algebra 
due to the dilaton.
\vspace{0.5cm}

In this study, we focus on the extension of the asymptotic symmetry algebra induced by symmetry breaking mechanisms, particularly
within the $sl(3,\mathbb{R})$ extension of the JT model. This allows for a detailed examination of how the algebraic structure 
is modified and what implications arise for the underlying dynamics. Previous studies have shown that although a Virasoro algebra
can be recovered in two-dimensional dilaton gravity~\cite{Grumiller:2013swa}, non-integrability of the associated charges 
posed limitations for establishing a consistent AdS$_2$/CFT$_1$ correspondence. This issue was later addressed by incorporating 
dilaton-dependent charges and relaxing the requirement of Casimir conservation~\cite{Grumiller:2017qao}, thereby enabling the
emergence of a full Virasoro algebra and enhancing the CFT interpretation. These developments underscore the importance of extended 
boundary conditions in AdS$_2$ gravity and open new directions for analyzing its holographic properties.
\vspace{0.5cm}
 
In conclusion, this study aims to thoroughly examine the interactions of two dimensional dilaton gravity 
with boundary conditions and symmetries, filling gaps in the existing literature. The potential and 
symmetric properties of JT gravity offer new perspectives on how these theories can be utilized to 
understand holographic dualities. In this context, two dimensional gravity theories have become one 
of the significant areas of theoretical physics, and this research will illuminate future studies.
\vspace{0.5cm}

\textbf{Related Work and Context:}
The relation between two-dimensional dilaton gravity and SYK-like models has received renewed attention in recent years. 
In particular, the work~\cite{Momeni:2024ygv} proposes a deformed SYK model as a holographic dual to generalized JT gravity, 
suggesting that modified boundary terms and higher-spin structures can encode new aspects of the dual theory. 
\vspace{0.5cm}

In parallel, studies such as~  \cite{Momeni:2020tyt,Momeni:2020zkx,Momeni:2021jhk} have investigated higher-spin extensions 
of AdS$_2$ gravity, emphasizing the role of extended symmetry algebras—especially $\mathcal{W}$-algebras—in organizing asymptotic
dynamics. Our approach aligns with these developments by realizing an $sl(3,\mathbb{R})$ extension of JT gravity within 
a BF-theoretic framework, which accommodates symmetry breaking and coadjoint orbit structures. This provides a concrete
realization of generalized boundary dynamics that fits naturally within the broader higher-spin gravity landscape.
\vspace{0.5cm}

This paper is organized as follows: Section~\ref{sec:secQ} discusses the fundamental aspects of the quantum \( \mathcal{W}_3 \) 
algebra, including its conformal structure and its extended version with specific conformal spin properties. Section~\ref{sec:sec21} 
explores the \( sl(2,\mathbb{R}) \) holographic dictionary and examines JT dilaton gravity as a two--dimensional 
BF theory. Furthermore, connections to \( sl(2,\mathbb{R}) \) BF theory are established, and various boundary 
conditions, particularly affine and conformal boundary conditions, are analyzed in detail. Section~\ref{sec:spin3} focuses on 
\( sl(3,\mathbb{R}) \) higher-spin dilaton gravity, where the effects of higher--spin theories on boundary conditions 
are explored, and the roles of affine and conformal boundary conditions in this context are evaluated. Section~\ref{sec:conclusion} 
provides a comprehensive discussion of the results, emphasizing both theoretical and practical implications. 
Finally, Section~\ref{sec:ack} includes acknowledgments to contributors and supporting institutions.
\section{The quantum $\cW_3$--algebra}
\label{sec:secQ}

We begin with a brief review of the fundamental aspects of the quantum $\cW_3$--algebra.\,This section does not aim to provide an exhaustive discussion;\,rather,\,it highlights the essential properties that are most pertinent to our study.\,Originally introduced by A.B. Zamolodchikov \cite{Zamolodchikov:1985wn}, the quantum $\cW_3$--algebra extends the Virasoro algebra in a non--linear fashion, naturally incorporating higher--spin symmetries.

\subsection{Conformal structure of the quantum $\cW_3$--algebra}\label{sec21}

The structure of the system involves two fundamental fields: a single spin--$3$ field, denoted as $W(z)$, and a simpler spin--$2$ field, represented by $T(z)$. These fields follow the conventional mode expansions:
$T(z)=  \sum_n L_n z^{-n-2}~,\,$ and  $W(z)= \sum_n W_n z^{-n-3}~$. The non--trivial operator product expansions  that characterize the quantum $\mathcal{W}_3$--algebra adhere to the standard non--linear structure,

\begin{align}
T(z_1) T(z_2) &\sim \frac{c}{2} \frac{1}{z_{12}^4} + \frac{2 T}{z_{12}^2} + \frac{\partial T}{z_{12}}, \\[10pt]
T(z_1) W(z_2) &\sim \frac{3 W}{z_{12}^2} + \frac{\partial W}{z_{12}},\\ 
W(z_1) W(z_2) &\sim \frac{c}{3} \frac{1}{z_{12}^6} + \frac{2 T}{z_{12}^4} + \frac{\partial T}{z_{12}^3} \notag \\[10pt]
&\quad + \frac{1}{z_{12}^2} \left(2b^2 \Lambda + \frac{3}{10} \partial^2 T \right) 
 + \frac{1}{z_{12}} \left( b^2 \partial \Lambda + \frac{1}{15} \partial^3 T \right).
\end{align}

where $z_{12}=z_1-z_2$,\,$\Lambda= :TT:-\frac{3}{10} \partial^2 T$,\,$b^2=\frac{16}{5c+22}$ and $c$ is the central charge.

\subsection{Extended $\cW_3$--algebra with conformal spin--$(\Delta,1-\Delta)$  }
In this section, we extend the quantum $\cW_3$--algebra with conformal spins $\Delta = 2$ and $\Delta = 3$, as introduced in the previous subsection, 
to construct the extended $\cW_3$--algebra. This is achieved by incorporating two additional fields, referred to as the dilaton fields, 
$X$ and $Y$, which possess conformal spins $1 - \Delta = -1$ and $-2$, respectively. All having the standard mode expansions
$X(z)=  \sum_n X_n z^{-n+1}~,\,$ and  $Y(z)= \sum_n Y_n z^{-n+2}~$. Using Thielemans’ Mathematica package \cite{Thielemans:1991uw}, 
one can show that the resulting extended $\cW_3$-- algebra with non--zero operator product expansions is obtained as follows:
\begin{align}
T(z_1) T(z_2) &\sim \frac{c}{2} \frac{1}{z_{12}^4} + \frac{2 T(z_2)}{z_{12}^2} + \frac{\partial T(z_2)}{z_{12}}, \label{Q1} \\[10pt]
T(z_1) W(z_2) &\sim \frac{3 W(z_2)}{z_{12}^2} + \frac{\partial W(z_2)}{z_{12}}, \\[10pt]
W(z_1) W(z_2) &\sim \frac{c}{3} \frac{1}{z_{12}^6} + \frac{2 T(z_2)}{z_{12}^4} + \frac{\partial T(z_2)}{z_{12}^3} \nonumber \\
&\quad + \frac{1}{z_{12}^2} \left(2b^2 \Lambda(z_2) + \frac{3}{10} \partial^2 T(z_2) \right)
+ \frac{1}{z_{12}} \left( b^2 \partial \Lambda(z_2) + \frac{1}{15} \partial^3 T(z_2) \right), \\[10pt]
T(z_1) X(z_2) &\sim - \frac{X(z_2)}{z_{12}^2} + \frac{\partial X(z_2)}{z_{12}}, \\[10pt]
T(z_1) Y(z_2) &\sim - \frac{2 Y(z_2)}{z_{12}^2} + \frac{\partial Y(z_2)}{z_{12}},
\end{align}

\begin{align}
W(z_1) X(z_2) &\sim - \frac{2 Y(z_2)}{5 z_{12}^{4}} + \frac{\partial Y(z_2)}{5 z_{12}^{3}} \nonumber \\[10pt]
&\quad -\frac{1}{z_{12}^2} \left(\frac{32 T Y}{5(c-28)} + \frac{(c+4)\partial^2 Y}{10(c-28)} \right) \nonumber \\[10pt]
&\quad + \frac{1}{z_{12}} \left( \frac{32(c+2) T \partial Y }{5(c-28)^2} + \frac{192 \partial T Y }{(c-28)^2} +
\frac{(c^2+24 c -16)\partial^3 Y}{15(c-28)^2} \right), \\[10pt]
W(z_1) Y(z_2) &\sim - \frac{X(z_2)}{z_{12}^2} + \frac{2 \partial X(z_2)}{z_{12}}. \label{Q2}
\end{align}
This extended quantum  $\mathcal{W}_3$--algebra version will be compared with the classical version that will be calculated 
at the end of Section~\ref{sec:spin3}.
\section{The $sl(2,\mathbb{R})$ holographic dictionary}\label{sec:sec21}

We analyze in this section the framework of two--dimensional dilaton gravity formulated on $AdS_2$. Our discussion begins with a brief introduction to the fundamental aspects of dilaton gravity, emphasizing elements crucial for establishing the holographic dictionary. We then proceed with a systematic derivation of the asymptotic symmetry algebra in the context of $sl(2,\mathbb{R})$. Afterward, we delve into the role of external sources and the associated holographic Ward identities. These identities serve as essential tools for investigating higher--spin black hole configurations.

\subsection{JT dilaton gravity as a BF theory in two dimensions}\label{sec2}

This section presents a brief summary of $AdS_2$ higher--spin gravity within the framework of dilaton gravity. We adopt the gauge approach to dilaton gravity, treating it as a non--abelian model. In particular, this formalism is utilized to explore $AdS_2$ gravity based on the $sl(2,\mathbb{R})$ algebra structure.

\subsection{Connection to $sl(2,\mathbb{R})$ BF theory}

Similar to the technical advantages offered by Chern–-Simons theory in the three dimensional
 framework \cite{Achucarro:1986uwr, Witten:1988hc}, alternative formulations in two dimensional 
 gravity systems provide significant simplifications. In this study, we will briefly outline the 
 key features of these approaches, focusing specifically on the JT model. Therefore, in two dimensions,
  the dilaton gravity action  with a negative cosmological constant can be equivalently formulated 
  over a spacetime manifold $\cM$:

\begin{equation}
    S[\cX,\cA] = \frac{k}{4\pi}\int_{\cM} \mathfrak{tr}[\cX \cF] +S_{\rm bdy} ,
\end{equation}

where $\mathcal{A}$ denotes a 1--form connection associated with the field strength 
$\cF= \mathrm{d}\cA + \cA\wedge \cA$, and $k$ represents a coupling constant. The dilaton field $\cX$ is an algebra--valued scalar. The manifold $\cM$ is assumed to have the topology of $S^1$, with the radial coordinate extending over the range $0\leq\rho<\infty$. Additionally, the Euclidean time coordinate $y$ is periodically identified as $y \sim y +\beta$, where $\beta$ corresponds to the inverse temperature. The term $S_{\rm bdy}$ is introduced to ensure a well--defined action principle, enforcing appropriate boundary conditions at $\partial \mathcal{M}$, which corresponds to the asymptotic boundary at infinite radius. The notation used throughout the paper, including gauge field components, dilaton variables, and boundary charges, follows standard conventions. For clarity, we summarize the main symbols and their roles in Appendix ~\ref{sec:appa}.\\

The fields $\cA$ and $\cX$ take values in the gauge algebra $sl(2,\mathbb{R})$. The trace operation $\mathfrak{tr}$ provides a metric structure for the generators of the $sl(2,\mathbb{R})$ Lie algebra. In order to describe the dilaton gravity system, one employs the generators $\Lt_i$ ($i=0,\pm1$), which satisfy the algebraic structure:

\begin{eqnarray}
    \left[\Lt_i, \Lt_j\right] &=& (i-j) \Lt_{i+j}.
\end{eqnarray}
We choose a matrix representation for its generators,
\begin{equation}
\Lt_{-1}=\begin{pmatrix}0 & 0\\
1 & 0
\end{pmatrix}\quad,\quad \Lt_{0}=\frac{1}{2}\begin{pmatrix}-1 & 0\\
0 & 1
\end{pmatrix}\quad,\quad \Lt_{1}=\begin{pmatrix}0 & -1\\
0 & 0
\end{pmatrix}\;.\label{PTT-sl(2,R)-MR}
\end{equation}
Then, the only non-zero components of the invariant bilinear form are given by 
$\mathfrak{tr}( \Lt_{\mp1}\Lt_{\pm1})=-2 \mathfrak{tr}(\Lt_{0}\Lt_{0})=-1$. The action is invariant under the gauge transformations, 
\begin{equation}\label{gauge transformation}
\delta_{\lambda} \cA =d\lambda + \left[\cA,\lambda \right],\qquad \qquad \delta_{\lambda} \cX =[\cX,\lambda],
\end{equation}
where $\lambda$ is also $sl(2,\mathbb{R})$ Lie algebra--valued gauge parameter. 

These gauge transformations characterize the residual symmetries of the theory that preserve the asymptotic form of the gauge fields, and also
encode how the gauge parameters generate boundary symmetries within the BF formulation. These structures form the basis for identifying 
the asymptotic symmetry algebra, which in the $sl(2,\mathbb{R})$ case corresponds to Virasoro, and in the $sl(3,\mathbb{R})$ case extends to the $\mathcal{W}_3$ algebra. 
The inclusion of the dilaton field plays a crucial role in controlling which symmetries are preserved or broken at the boundary.
The equations of motion are given by
\begin{equation} \label{eom}
\mathcal{F} =d\cA+\cA\wedge \cA=0,\qquad \qquad d \cX + \left[\cA,\cX\right]=0.
\end{equation}
  More importantly, these gauge transformations determine the residual symmetries that survive after fixing the radial gauge near the asymptotic boundary of AdS$_2$. As such, they form the foundation for identifying the asymptotic symmetry algebra (ASA). The presence of the dilaton field $X$ plays a crucial role here: it interacts with the gauge connection at the boundary and imposes dynamical constraints that restrict the allowed gauge transformations. This leads to a partial breaking of the full infinite-dimensional affine symmetry algebra, such as: $\mathfrak{sl}(2,\mathbb{R})_k$ or $\mathcal{W}_3$, down to their respective finite-dimensional subalgebras: $\mathfrak{sl}(2,\mathbb{R})$ or $\mathfrak{sl}(3,\mathbb{R})$. In this sense, equations \eqref{gauge transformation} and \eqref{eom} do not merely define local gauge redundancies, but also encode the physical mechanism by which symmetry breaking occurs in the presence of a dilaton.
\vspace{0.5cm}

By employing the radial gauge, the connections in an asymptotically \( AdS_2 \) spacetime can be expressed as
\begin{eqnarray}
\label{ads31}
    \cA &=& b^{-1} a\left(t\right) b + b^{-1} \mathrm{d}b, \quad
    \cX =b^{-1} x\left(t\right) b,
\end{eqnarray}
where the group element \( b(\rho) \), which does not depend on the state, takes the form
\begin{eqnarray}
\label{gr}
    b(\rho) = e^{\rho \Lt_{0}},
\end{eqnarray}
This representation facilitates a broader class of metrics incorporating all $sl(2,\mathbb{R})$ charges. Notably, as long as \( \delta b = 0 \), the specific choice of \( b \) does not influence asymptotic symmetries. This flexibility permits a more comprehensive metric formulation, necessitating boundary conditions that maintain this generality in the gravitational setting. Moreover, within the radial gauge framework, the connection \( a(t) \) is a field valued in the $sl(2,\mathbb{R})$ Lie algebra, remaining independent of the radial coordinate:
 $a\left(t\right) =a_{t}\left(t\right)\mathrm{d}t$.
 
Our analysis focuses on the asymptotic boundary conditions of the $sl(2,\mathbb{R})$ dilaton theory in the affine scenario. We demonstrate how the methodology developed in 
\cite{Grumiller:2016pqb,Ozer:2017dwk,Ozer:2019nkv,Ozer:2021wtx,Ozer:2024ovo} 
can be utilized to examine the asymptotic symmetry algebra. Based on these findings, the most general solution to the equations governing an asymptotically $AdS_2$ spacetime can be written in the following general metric form:
\begin{gather}\label{AdSLineElement}
	\extd s^2 = \extd\rho^2 + 2\,\mathcal{L}^{0} \extd\rho \extd\varphi + \Big(\left(e^{\rho}\mathcal{L}^{+}
    -e^{-\rho}\mathcal{L}^{-}\right)^{2}+\left(\mathcal{L}^{0}\right)^{2}\Big)\,\extd\varphi^2 ~,
\end{gather}
which is s reminiscent of its $AdS_3$ version \cite{Grumiller:2016pqb,Ozer:2017dwk,Ozer:2019nkv,Ozer:2021wtx,Ozer:2024ovo}. The dilaton is also,
\begin{gather}\label{AdSDilaton}
	X = e^{\rho}\,\mathcal{X}^{+} + e^{-\rho}\,\mathcal{X}^{-} ~.
\end{gather}
Therefore, it is crucial to define affine  $sl(2,\mathbb{R})$ gravity boundary conditions that preserve this form of the metric.
\subsection{Affine boundary conditions}\label{osp21}

The goal of this section is to formulate $sl(2,\mathbb{R})$ higher--spin $AdS_2$ dilaton gravity at the affine boundary. We carry out our computations to clarify the asymptotic symmetry algebra under
the loosest set of boundary conditions. We begin by introducing the $sl(2,\mathbb{R})$ Lie algebra--valued $a_t$ component of the gauge connection as:
\begin{eqnarray}\label{bouncond999}
a_t &=&\alpha_i\mathcal{L}^i \Lt_{i}
\end{eqnarray}
where the coefficients are given by $\alpha_{0}=-2\alpha_{\pm1}=\frac{4}{k}$. Here, we identify three state--dependent functions, denoted as $\mathcal{L}^i$, which are commonly referred to as charges. The dilaton field $x$ takes the form:
\begin{eqnarray}\label{bouncond888}
x &=&\cX^i \Lt_{i}.
\end{eqnarray}
This setup similarly leads to three state-dependent functions, represented as $\cX^i$. Our objective is to extract the asymptotic symmetry algebra within affine boundary conditions using a canonical analysis. To this end, we examine all gauge transformations \eqref{gauge transformation} that preserve the affine boundary conditions.

At this point, it is useful to express the gauge parameter using the $sl(2,\mathbb{R})$ Lie algebra basis:
\begin{equation}\label{boundarycond2}
    \lambda = b^{-1}\left[\epsilon^i \Lt_{i} \right]b.
\end{equation}
Here, the gauge parameter involves three bosonic variables $\epsilon^i$, which are arbitrary functions of the boundary coordinates. Next, we turn our attention to gauge parameters satisfying \eqref{gauge transformation}. The corresponding infinitesimal gauge transformations are given by:

 \begin{align} 
       2\delta_\lambda\mathcal{L}^{0}	&= -\frac{k}{4}\partial_t\epsilon^{0} - \mathcal{L}^{+1}\epsilon^{-1} + \mathcal{L}^{-1}\epsilon^{+1}, \label{transf21} \\ 
    \delta_\lambda\mathcal{L}^{\pm 1}	&= \frac{k}{2}\partial_t\epsilon^{\mp 1} + \mathcal{L}^{0}\epsilon^{\mp 1} \mp \mathcal{L}^{\mp 1}\epsilon^{0}, \label{transf22} \\ 
    \delta_\lambda\mathcal{X}^{0}	&= 2\big(\mathcal{X}^{+1}\epsilon^{-1} - \mathcal{X}^{-1}\epsilon^{+1}\big), \label{transf23} \\ 
    \delta_\lambda\mathcal{X}^{\pm 1}	&=\mp \mathcal{X}^{0}\epsilon^{\pm 1} \pm \mathcal{X}^{\pm 1}\epsilon^{0}. \label{transf24}
\end{align}
 
As a concluding step, we introduce the canonical boundary charge $\mathcal{Q}_a[\lambda]$, which is responsible for generating the transformations described in Eqs.~\eqref{transf21}–\eqref{transf22}. Likewise, we define the canonical boundary charge $\mathcal{Q}_x[\lambda]$, which governs the transformations given in Eqs.~\eqref{transf23}–\eqref{transf24} for the dilaton field.
The infinitesimal variation of these charges \cite{Banados:1994tn}, which gives rise to the asymptotic symmetry algebra, takes the form:

\begin{equation}\label{Qvar}
    \delta_\lambda \mathcal{Q}_a = \frac{k}{2\pi} \int \mathrm{d}t\; \mathfrak{tr} \left(\lambda \delta a_{t} \right), \quad
    \delta_\lambda \mathcal{Q}_x = \frac{k}{2\pi} \int \mathrm{d}t\; \mathfrak{tr} \left(\lambda \delta x_{} \right).
\end{equation}

By integrating these variations functionally, we obtain the explicit expression for the canonical boundary charge:
\begin{equation}\label{boundaryco9}
    \mathcal{Q}_a[\lambda] = \int \mathrm{d}t\; \left( \mathcal{L}^{i} \epsilon^{-i} \right), \quad
    \mathcal{Q}_x[\lambda] = \int \mathrm{d}t\; \left( \mathcal{X}^{j} \epsilon^{-j} \right).
\end{equation}
Having established both the infinitesimal transformations and the corresponding boundary charge, we proceed to derive the asymptotic symmetry algebra by employing the conventional approach \cite{Blagojevic:2002du}. This structure is determined through the following relation:
\begin{equation}\label{Qvar2}
    \delta_{\lambda} \digamma = \{\digamma, \mathcal{Q}_{a,x}[\lambda]\}
\end{equation}
for any phase space functional $\digamma$. The asymptotic symmetry algebra is then generated by the charges $\mathcal{L}^i$ and $\mathcal{X}^i$. Ultimately, the operator product algebra can be expressed as follows:
\begin{eqnarray}\label{ope22}
    \mathcal{L}^{i}(\tau_1)\mathcal{L}^{j}(\tau_2) & \sim & \frac{\frac{k}{2}\eta^{ij}}{\tau_{12}^{2}} + \frac{(i-j)}{\tau_{12}} \mathcal{L}^{i+j}\\
    \mathcal{L}^{i}(\tau_1)\mathcal{X}^{j}(\tau_2) & \sim & \frac{(2i+j)}{\tau_{12}} \mathcal{X}^{i+j},\\
\end{eqnarray}
where $\tau_{12}=\tau_1-\tau_2$. The coefficients $\eta^{ij}=\mathfrak{tr}(\Lt_i\Lt_j)$ correspond to bilinear forms in the fundamental representation of the $sl(2,\mathbb{R})$ Lie algebra. An alternative way to present the operator product algebra in a more concise notation is:
\begin{eqnarray}\label{ope11}
    \mathfrak{\mathcal{\mathfrak{J}}}^{A}(\tau_1)\mathfrak{\mathcal{\mathfrak{J}}}^{B}(\tau_2) & \sim &
    \frac{\frac{k}{2}\eta^{AB}}{\tau_{12}^{2}} + \frac{\mathfrak{\mathcal{\mathfrak{f}}}^{AB}_{~~~C}
    \mathfrak{\mathcal{\mathfrak{J}}}^{C}}{\tau_{12}}. 
\end{eqnarray}
Here, $\eta^{AB}$ represents the trace matrix, while $\mathfrak{\mathcal{\mathfrak{f}}}^{AB}_{~~C}$ are the structure constants of the underlying algebra, where indices run as $(A,B=0,\pm1)$ and specifically $\mathfrak{\mathcal{\mathfrak{f}}}^{ij}_{~~i+j}=(i-j)$. Ultimately, for the most relaxed boundary conditions in $sl(2,\mathbb{R})$ dilaton gravity, the asymptotic symmetry algebra is identified as a single copy of the affine $sl(2,\mathbb{R})_k$ algebra.
\vspace{0.5cm}

The time--dependent behavior of the dilaton field at the affine boundary produces a structural effect that extends the classical ASA of the $sl(2,\mathbb{R})$ JT dilaton gravity. In the standard BF formulation, the ASA is derived solely from the gauge transformations of the $sl(2,\mathbb{R})$--valued connection field $a_t$. However, when the $sl(2,\mathbb{R})$ --valued dilaton field $x$ is included as a dynamical component, the preserved symmetry subspace becomes restricted, giving rise to a richer physical structure. This extended framework necessitates the incorporation into the symmetry algebra not only of the reparametrization modes, but also of new currents sourced by the dilaton.
\vspace{0.5cm}

As a result, the emerging symmetry structure requires the inclusion of abelian currents in addition to classical $sl(2,\mathbb{R})_k$  affine algebras. This indicates that the affine boundary dynamics of JT gravity cannot be fully described by the Schwarzian action alone, but instead calls for a multi-mode, extended action. Consequently, the dilaton becomes a determining element not only of the background geometry but also of the algebraic structure at the boundary.
\color{black}
\subsection{Conformal boundary conditions}\label{bhreduction1}
In this section, our objective is to examine the asymptotic symmetry algebra for the Brown--Henneaux boundary conditions. 
This gauge is chosen to explicitly capture the asymptotic symmetry structure and simplify the analysis of boundary dynamics. 
By separating the radial dependence, it isolates the asymptotic dynamics, facilitating the derivation of conserved quantities 
and boundary theories. This gauge plays a crucial role in preserving the asymptotic symmetry group and determining effective 
boundary actions, particularly in the context of holographic duality. We begin by enforcing the Drinfeld--Sokolov highest weight gauge condition on the $sl(2,\mathbb{R})$ Lie algebra--valued connection \eqref{bouncond999} to further restrict the coefficients. This reduction constrains the fields as follows:
\begin{eqnarray}
\mathcal{L}^0 = 0, \quad \mathcal{L}^{-1} = \mathcal{L}, \quad \alpha_{+1}\mathcal{L}^{+1} = 1.
\end{eqnarray}
where $\alpha_{-1} = \alpha$. It is worth mentioning that the conformal boundary conditions correspond to the well--established Brown--Henneaux boundary conditions formulated in \cite{Brown:1986nw} for $AdS_2$ gravity. Motivated by boundary conditions introduced in three-dimensional gravity \cite{Perez:2016vqo,Cardenas:2021vwo}, we propose that the gauge connection and dilaton take the following structure:
\begin{align}\label{bouncondhf}
a_t &= \Lt_{+1} + \alpha\mathcal{L} \Lt_{-1},\\
x &= \cX \Lt_{1}
 -\cX'\Lt_{0}
+ \bigg( \frac{\mathcal{L}\cX}{\alpha}
             +\frac{\cX ''}{2}
       \bigg)\Lt_{-1}.
\end{align}
Here, $\alpha$ serves as a scaling parameter, whose precise value will be determined later. We define two functionals: the charge $\mathcal{L}$ and the dilaton $\cX$. Upon implementing these constraints, we are positioned to derive the conformal asymptotic symmetry algebra. Based on the implications of the Drinfeld--Sokolov reduction, the gauge parameter $\lambda$ is governed by four independent functions, with $(\mathcal{\epsilon}\equiv\epsilon^{+1})$, taking the form:
\begin{eqnarray}\label{ebc08}
\lambda &=& b^{-1}\left[
  \epsilon \Lt_{1}
 -\epsilon'\Lt_{0}
+ \bigg( \frac{\mathcal{L}\epsilon}{\alpha}
             +\frac{\epsilon ''}{2}
       \bigg)\Lt_{-1}      
\right]b.
\end{eqnarray}

The dilaton field \(x\) is defined in the same form as the gauge parameter \(\lambda\) to ensure compatibility with the equations of motion 
and the asymptotic symmetry group. This structure allows the dilaton field to behave as a coadjoint element of the gauge theory and act as 
a stabilizer of the gauge symmetries. Furthermore, this choice ensures the well-posedness of the variational principle and the proper derivation 
of the conserved Casimir function \(C\). Consequently, defining \(x\) in the same form as \(\lambda\) is essential for the consistency of 
the theory, the preservation of the asymptotic symmetry group, and the formulation of boundary theories in the context of holographic duality.

By substituting the gauge parameter \(\lambda\) into the field transformation expressions \eqref{gauge transformation}, the infinitesimal gauge transformations take the following form:

\begin{eqnarray}
\delta_{\lambda}\mathcal{L} &=&
\frac{1}{2 \alpha}  \epsilon ^{'''}
+\epsilon  \mathcal{L}'+2 \mathcal{L} \epsilon', \label{ebc08666} \\
\delta_{\lambda}\cX &=&
\epsilon  \cX'- \epsilon' \cX. \label{ebc08667}
\end{eqnarray}

For the prescribed boundary conditions, the associated boundary charges $\mathcal{Q}_a[\lambda]$ and $\mathcal{Q}_x[\lambda]$ remain integrable, allowing us to interpret $\mathcal{L}$ and $\mathcal{X}$ as conserved charges. These transformations play a crucial role in understanding the structure of the asymptotic symmetry algebra \cite{Banados:1998pi}. 

Through an analysis of asymptotic symmetries, one can integrate the variation of the canonical boundary charges, represented by \(\delta_\lambda \mathcal{Q}_a\) in \eqref{ebc08666} and by \(\delta_\lambda \mathcal{Q}_x\) in \eqref{ebc08667} , leading to the expressions:

\begin{equation}\label{boundaryco11111333333}
    \mathcal{Q}_a[\lambda] = \int \mathrm{d}\varphi\; \left(\mathcal{L} \epsilon \right), \quad
    \mathcal{Q}_x[\lambda] = \int \mathrm{d}\varphi\; \left(\mathcal{X} \epsilon \right).
\end{equation}

These canonical boundary charges naturally facilitate an operator product algebra at the asymptotic boundary. Unlike two--dimensional CFT, a one--dimensional CFT scenario can be realized by effectively suppressing one coordinate, setting \(z_2\) to zero in \(z\), and identifying \(z_1\) with \(\tau\), or simply setting \(z = \tau\):
$z_1$ as $\tau$ or, more simply, by setting $z = \tau $:
\begin{eqnarray}\label{ope2}
  \mathcal{L}(\tau_1)\mathcal{L}(\tau_2) &\sim& \frac{{3k}}{{2\tau_{12}^{4}}} + \frac{2\mathcal{L}}{{\tau_{12}^{2}}} + \frac{\mathcal{L}'}{\tau_{12}},\\
   \cL(\tau_1)\cX(\tau_2) &\sim& - \frac{\cX}{{\tau_{12}^{2}}} + \frac{\cX'}{\tau_{12}}.
  \end{eqnarray}

From these relations, it follows that \(\cX\) transforms as a vector at the boundary, whereas \(\cL\) corresponds to a spin-two density. Throughout this work, the explicit dependence on \(\lambda\) has been removed, as it is no longer a necessity imposed by the boundary conditions. In essence, the residual symmetries associated with \(a_t\) and \(x\) correspond to diffeomorphisms on the circle, governed by the vector field \(\epsilon\). It is observed that the asymptotic symmetry algebra of \(sl(2,\mathbb{R})\) dilaton gravity under conformal boundary conditions is given by a single Virasoro algebra with a central charge \(c=3k\).\\
\vspace{0.5cm}

The conformal boundary behavior of the dilaton field introduces a structural modification that generically leads to an extension of the classical ASA in $sl(2,\mathbb{R})$ JT dilaton gravity as in the affine case.
In the conventional BF approach, the ASA is constructed exclusively from the gauge transformations associated with the $sl(2,\mathbb{R})$-valued connection field $a_t$. Once the $sl(2,\mathbb{R})$-valued dilaton $x$ is treated as an active dynamical field, the residual symmetry subalgebra becomes further constrained, resulting in a richer and more intricate physical framework.This generalized construction implies that the symmetry algebra must accommodate not only the reparametrization degrees of freedom, but also additional currents directly generated by the dilaton configuration.
\vspace{0.5cm}

Accordingly, the emerging boundary symmetry must contain abelian contributions that go beyond the classical Virasoro structure. This observation suggests that the Schwarzian action is insufficient to capture the full boundary dynamics in the presence of a nontrivial dilaton, and must be replaced by a multi--component, extended action principle. As a result, the dilaton field plays a dual role: it governs both the spacetime geometry and the underlying algebraic content of the boundary theory.
\section{$sl(3,\mathbb{R})$ higher--spin dilaton gravity}\label{sec:spin3}

Building upon the canonical analysis of the $sl(3,\mathbb{R})$ case, we introduce the extended higher--spin gravity framework rooted in the $sl(3,\mathbb{R})$ Lie algebra. The structure of this algebra consists of the conventional $sl(2,\mathbb{R})$ generators, denoted as $\Lt_i$, along with additional spin--3 generators $\Wt_m$ for indices ranging from $-2$ to $2$. The commutation relations among these elements are expressed as follows:
\begin{align}
[\Lt_i,\Lt_j]&=(i-j)\Lt_{i+j} \label{sl2com}\\
[\Lt_i,\Wt_m]&=(2i-m)\Wt_{i+m}\\
[\Wt_n,\Wt_m]&=\sigma(n-m)(2n^2-nm+2 m^2-8)\Lt_{n+m}
\end{align}
Additionally, the non-trivial components of the invariant bilinear forms are given by 
$\mathfrak{tr}( \Lt_{\mp1}\Lt_{\pm1})=-2 \mathfrak{tr}( \Lt_{0}\Lt_{0})=-1$ and 
$\mathfrak{tr}( \Wt_{\mp2}\Wt_{\pm2})=-4 \mathfrak{tr}( \Wt_{\mp1}\Wt_{\pm1})=6 \mathfrak{tr}( \Wt_{0}\Wt_{0})=-48 \sigma$. 
\subsection{Affine boundary conditions}

The aim of this section is to develop an extended higher--spin $AdS_2$ framework by formulating it as an $sl(3,\mathbb{R})$ dilaton theory defined on the affine boundary. 
We carry out our analysis to determine the asymptotic symmetry algebra under the broadest possible boundary conditions. As discussed in the preceding section, we adopt the principal embedding of $sl(2,\mathbb{R})$ into $sl(3,\mathbb{R})$ as a subalgebra. This allows us to prescribe affine boundary conditions for asymptotically $AdS_2$ spacetimes. To facilitate this construction, we introduce the gauge connection in the following form:
\begin{align}\label{bouncondaf}
    a_t = \alpha_i\mathcal{L}^i \Lt_{i}
    +\beta_i\mathcal{W}^i \Wt_{i}
   \end{align}
   
where the scaling relations among the parameters are given by $\alpha_{0}=-2\alpha_{\pm1}$ and $\beta_{0}=-\frac{3}{2}\beta_{\pm1}=6 \beta_{\pm2}$. As a result, there exist eight dynamical functions: $\mathcal{L}^i$ and $\mathcal{W}^i$, which serve as the charges. The dilaton field $x$ takes the form:
\begin{eqnarray}\label{bouncond888}
x =\cX^i \Lt_{i}+\cY^i \Wt_{i}.
\end{eqnarray}
In this setting, we obtain eight additional state-dependent functions, $\cX^i$ and $\cY^i$. Our objective is to derive the asymptotic symmetry algebra corresponding to the affine boundary conditions through a canonical analysis. To accomplish this, we systematically examine all gauge transformations that preserve these boundary conditions. 

At this point, it is convenient to express the gauge parameter using the basis of the $sl(3,\mathbb{R})$ Lie algebra:
\begin{equation}\label{boundarycond2}
    \lambda = b^{-1}\left[\epsilon^i \Lt_{i}+\eta^i \Wt_{i} \right]b.
\end{equation}
Here, the gauge parameter comprises eight bosonic variables, $\epsilon^i$ and $\eta^i$, which are arbitrary functions of the boundary coordinates. Next, we proceed to analyze the gauge parameters satisfying \eqref{gauge transformation} and determine the corresponding infinitesimal gauge transformations.

\begin{align}
\delta_\lambda \cL^{\pm 1} =&-\frac{2 \partial_t\epsilon^{\pm 1}}{\alpha_0} \pm \cL^{\pm 1} \epsilon^0 \pm 2\cL^0 \epsilon^{\pm 1}\nonumber \\
                          &\pm \cW^{\pm 2} \eta^{\mp 1} \pm 2\cW^{\pm 1} \eta^0 \pm 3\cW^0 \eta^{\pm 1} \pm 4\cW^{\mp 1} \eta^{\pm 2},\label{transf41}  \\
\delta_\lambda \cL^0 =& \frac{\partial_t\epsilon^0}{\alpha_0}-\cL^{+1} \epsilon^{-1} + \cL^{-1} \epsilon^{+1}\nonumber \\
& - 2\cW^2 \eta^{-2} - \cW^{+1} \eta^{-1} + \cW^{-1} \eta^{+1} + 2\cW^{-2} \eta^2, \label{transf42}  \\
\delta_\lambda \cW^{\pm 2} =& - \frac{24 \sigma \partial_t\eta^{\pm 2}}{\alpha_0}\pm 2(\cW^{\pm 2} \epsilon^0 \pm 2\cW^{\pm 1} \epsilon^{\pm 1})\nonumber \\
& \pm 12 \sigma(\cL^{\pm 1} \eta^{\pm 1} \pm 4\cL^0 \eta^{\pm 2}), \label{transf43}\\
\delta_\lambda \cW^{\pm 1} =&  \frac{6 \sigma \partial_t\eta^{\pm 1}}{\alpha_0}\mp \cW^{\pm 2} \epsilon^{\mp 1} \pm \cW^{\pm 1} \epsilon^0 \pm 3\cW^0 \epsilon^{\pm 1}\nonumber \\
& \mp 6 \sigma(\cL^{\pm 1} \eta^0 \mp \cL^0 \eta^{\pm 1} \pm 2\cL^{-1} \eta^{\pm 2}),\label{transf44}  \\
\delta_\lambda \cW^0 =& - \frac{4 \sigma \partial_t\eta^0}{\alpha_0}-2(\cW^{+1} \epsilon^{-1} - \cW^{-1} \epsilon^{+1})\nonumber \\
& + 6 \sigma(\cL^{+1} \eta^{-1} - \cL^{-1} \eta^{+1}),\label{transf45}\\ 
\delta_\lambda \cX^{\pm 1} &= \mp \epsilon^{+1} \cX^0 \pm \epsilon^0 \cX^{+1}\nonumber\\
& \mp 6 \sigma(2 \eta^{\pm 2} \cY^{\mp 1} \mp \eta^{\pm 1} \cY^0 \pm \eta^0 \cY^{\pm 1} \mp 2 \eta^{\mp 1} \cY^{\pm 2}),\label{transf46}  \\
\delta_\lambda \cX^0 &= 2(\epsilon^{-1} \cX^{+1} - \epsilon^{+1} \cX^{-1})\nonumber \\
& - 6 \sigma(8 \eta^2 \cY^{-2} - \eta^{+1} \cY^{-1} + \eta^{-1} \cY^{+1} - 8 \eta^{-2} \cY^2),\label{transf47}  \\
\delta_\lambda \cY^{\pm 2} &=\mp 2 \eta^{\pm 2} \cX^0 \pm \eta^{\pm 1} \cX^{\pm 1} \mp \epsilon^{\pm 1} \cY^{\pm 1} \pm 2 \epsilon^0 \cY^{\pm 2}\nonumber  
\end{align} 
\begin{align}
\delta_\lambda \cY^{\pm 1} &= \pm 2 \eta^0 \cX^{\pm 1} \mp 4 \eta^{\pm 2} \cX^{\mp 1} \mp \eta^{\pm 1} \cX^0,\label{transf48} \\
& \mp 2 \epsilon^{\pm 1} \cY^0 \pm \epsilon^0 \cY^{\pm 1} \pm 4 \epsilon^{\mp 1} \cY^{\pm 2}\nonumber  \\
\delta_\lambda \cY^0 &= 3(\eta^{-1} \cX^{+1} - \eta^{+1} \cX^{-1} - \epsilon^{+1} \cY^{-1} + \epsilon^{-1} \cY^{+1}).\label{transf49} 
\end{align}
As a concluding step, we define the canonical boundary charge $\mathcal{Q}_{a,x}[\lambda]$ responsible for generating the transformations given in \eqref{transf41} - \eqref{transf49}. The variation of these charges, which governs the asymptotic symmetry algebra, is expressed as \cite{Banados:1994tn}

\begin{equation}\label{Qvar}
    \delta_\lambda \mathcal{Q}_a = \frac{k}{2\pi} \int \mathrm{d}t\; \mathfrak{tr} \left(\lambda \delta a_{t} \right), \quad
    \delta_\lambda \mathcal{Q}_x = \frac{k}{2\pi} \int \mathrm{d}t\; \mathfrak{tr} \left(\lambda \delta x_{} \right).
\end{equation}
Integrating this variation functionally yields the explicit form of the canonical boundary charges:
\begin{equation}\label{boundaryco9}
    \mathcal{Q}_a[\lambda] = \int \mathrm{d}t\; \left( \mathcal{L}^{i} \epsilon^{-i}+\mathcal{W}^{i}\eta^{-i} \right), \quad
    \mathcal{Q}_x[\lambda] = \int \mathrm{d}t\; \left( \mathcal{X}^{i} \epsilon^{-i}+\mathcal{Y}^{i}\eta^{-i} \right).
\end{equation}

Having established both the infinitesimal transformations and the canonical boundary charge, we are now equipped to derive the asymptotic symmetry algebra via the conventional approach \cite{Blagojevic:2002du}. This is achieved using the fundamental relation:
\begin{equation}\label{Qvar2}
    \delta_{\lambda} \digamma = \{\digamma, \mathcal{Q}_{a,x}[\lambda]\}
\end{equation}
for an arbitrary phase space functional $\digamma$. Here, the charges $\mathcal{L}^i$ , $\mathcal{W}^i$ and related dilatons $\mathcal{X}^i$ , $\mathcal{Y}^i$ serve as the generators of the asymptotic symmetry algebra. Ultimately, the operator product algebra can be expressed as follows:

\begin{eqnarray}\label{ope22}
\mathcal{L}^{i}(\tau_1)\mathcal{L}^{j}(\tau_2) & \sim & \frac{\frac{k}{2}\eta^{ij}}{\tau_{12}^{2}} + \frac{(i-j)}{\tau_{12}} \mathcal{L}^{i+j},\\
\mathcal{L}^{i}(\tau_1)\mathcal{W}^{j}(\tau_2) & \sim &                                              \frac{(2i-j)}{\tau_{12}} \mathcal{W}^{i+j},\\
\mathcal{W}^{i}(\tau_1)\mathcal{W}^{j}(\tau_2) & \sim & \frac{\frac{k}{2}\theta^{ij}}{\tau_{12}^{2}} + \sigma\frac{ (i - j) (2 i^2 - i j + 2 j^2 - 8)}{\tau_{12}} \mathcal{L}^{i+j},\\
\mathcal{L}^{i}(\tau_1)\mathcal{X}^{j}(\tau_2) & \sim & \frac{(2i+j)}{\tau_{12}} \mathcal{X}^{i+j},\\
\mathcal{L}^{i}(\tau_1)\mathcal{Y}^{j}(\tau_2) & \sim & \frac{(3i+j)}{\tau_{12}} \mathcal{Y}^{i+j},\\
\mathcal{W}^{i}(\tau_1)\mathcal{X}^{j}(\tau_2) & \sim &  \sigma\frac{ ( -16 i + 10 i^3 - 6 j - a4 j + 15 i^2 j + 9 i j^2 + a4 j^3)}{\tau_{12}} \mathcal{Y}^{i+j},\\
\mathcal{W}^{i}(\tau_1)\mathcal{Y}^{j}(\tau_2) & \sim &  \frac{ (3i+2 j)}{\tau_{12}} \mathcal{Y}^{i+j}.
\end{eqnarray}
where $\tau_{12}=\tau_1-\tau_2$. Here, $\eta^{ij}=\mathfrak{tr}(\Lt_i\Lt_j)$ and $\theta^{ij}=\mathfrak{tr}(\Wt_i\Wt_j)$ represent the bilinear forms in the fundamental representation of the $sl(3,\mathbb{R})$ Lie algebra. The operator product algebra can be rewritten in a more concise form:
\begin{eqnarray}\label{ope11}
    \mathfrak{\mathcal{\mathfrak{J}}}^{A}(\tau_1)\mathfrak{\mathcal{\mathfrak{J}}}^{B}(\tau_2) & \sim &
    \frac{\frac{k}{2}\eta^{AB}}{\tau_{12}^{2}} + \frac{\mathfrak{\mathcal{\mathfrak{f}}}^{AB}_{~~~C}
    \mathfrak{\mathcal{\mathfrak{J}}}^{C}}{\tau_{12}}. 
\end{eqnarray}
Here, $\eta^{AB}$ denotes the trace matrix, while $\mathfrak{\mathcal{\mathfrak{f}}}^{AB}_{~~C}$ correspond to the structure constants of the respective algebra, with $(A,B=0,\pm1,\pm2)$, satisfying the relation $\mathfrak{\mathcal{\mathfrak{f}}}^{ij}_{~~i+j}=(i-j)$. Finally, for  the loosest set of boundary conditions, the asymptotic symmetry algebra of $sl(3,\mathbb{R})$ dilaton gravity is characterized by a single copy of the affine $sl(3,\mathbb{R})_k$ algebra.
\vspace{0.5cm}

In $sl(3,\mathbb{R})$ JT dilaton gravity with affine boundary conditions, similar to the case of $sl(2,\mathbb{R})$ JT dilaton gravity under analogous constraints, the asymptotic symmetry algebra at the boundary is initially given by the infinite-dimensional $sl(3,\mathbb{R})_k$ algebra. However, the presence of the time-dependent dilaton field reshapes the asymptotic algebraic structure; as a result, even the expected preservation of the $SL(3,\mathbb{R})$ subgroup within the full $sl(3,\mathbb{R})_k$ symmetry is dynamically reduced, and the remaining infinite-dimensional symmetry components are broken. The dilaton acts as a stabilizer for the $sl(3,\mathbb{R})$ gauge connection and interacts with the gauge connection through boundary fluctuations, preventing the complete preservation of the infinite-dimensional $sl(3,\mathbb{R})_k$ symmetry. This reduction in symmetry is attributed to the dilaton field, which influences the free boundary fluctuations. Consequently, the symmetries associated with $a_t$, $x$, and $y$ correspond to diffeomorphisms on the circle generated by a vector field $\epsilon^{i}$ and $\eta^{i}$.

\subsection{Conformal boundary conditions}\label{bhreduction2}
In this section, we aim to analyze the asymptotic symmetry algebra associated with the Brown--Henneaux boundary conditions. 
To achieve this, we begin by enforcing the Drinfeld--Sokolov highest weight gauge condition on the $sl(3,\mathbb{R})$ Lie algebra--valued 
connection (\ref{bouncondaf}), thereby further restricting the coefficient structure. Consequently, the Drinfeld--Sokolov reduction leads to the following field constraints:
\begin{eqnarray}
\mathcal{L}^0 = 0, \quad \mathcal{L}^{-1} = \mathcal{L}, \quad \alpha_{+1}\mathcal{L}^{+1} = 1\\
\cW^{\pm1}=\cW^{0}, \quad \cW^{-1}=\cW, \quad \beta_{+2}\cW^{+2}=1..
\end{eqnarray}
and introduces scaling parameters $\alpha_{-1}=\alpha$ and $\beta_{-2}=\beta$. It is important to highlight that the conformal boundary conditions 
correspond to the well--established Brown--Henneaux boundary conditions originally formulated in \cite{Brown:1986nw} for $AdS_3$ gravity. 
This approach takes inspiration from boundary conditions previously proposed in the context of three-dimensional gravity \cite{Perez:2016vqo,Cardenas:2021vwo}.
Consequently, we propose the gauge connection and dilaton takes the form:
\begin{align}\label{bouncondhf}
a_t =&\Lt_{+1} + \alpha\mathcal{L} \Lt_{-1} +\beta\mathcal{W} \Wt_{-2},\\
x=&\cX \Lt_1  -\cX'\Lt_0 + \cY \Wt_2 -\cY' \Wt_1\nonumber\\
&+\left(\alpha \mathcal{L} \cX +\frac{\cX''}{2}+ 24 \beta  \sigma \mathcal{W} \cY+\right)\Lt_{-1}\nonumber\\
&+\Bigg(\beta  \mathcal{W} \cX+\frac{7}{12} \alpha  \cY' \mathcal{L}'+\frac{2}{3} \alpha  \mathcal{L} \cY''\nonumber\\
&+\frac{\cY^{(4)}}{24}+\alpha ^2 Y    \mathcal{L}^2+\frac{1}{6} \alpha \cY \mathcal{L}''\Bigg)\Wt_{-2}\\
&+\left(-\frac{5}{3} \alpha  \mathcal{L} \cY'-\frac{\cY^{(3)}}{6}-\frac{2}{3} \alpha  \cY \mathcal{L}'\right)\Wt_{-1}\nonumber\\
&+\left(\frac{\cY''}{2}+2 \alpha  \cY \mathcal{L}\right)\Wt_0.\nonumber
\end{align}
After carrying out these procedures, we arrive at the threshold of deriving the conformal asymptotic symmetry algebra. Based on the implications of the Drinfeld--Sokolov reduction, the gauge parameter $\lambda$ is characterized by only four independent functions, denoted as $\mathcal{\epsilon}\equiv\epsilon^{+1}$ and $\mathcal{\eta}\equiv\eta^{+2}$, which are explicitly given by:
\begin{align}\label{ebc08}
\lambda =
&b^{-1}\Bigg[\epsilon \Lt_1  -\epsilon'\Lt_0 + \eta \Wt_2 -\eta' \Wt_1\nonumber\\
&+\left(\alpha \mathcal{L} \epsilon +\frac{\epsilon''}{2}+ 24 \beta  \sigma \mathcal{W} \eta+\right)\Lt_{-1}\nonumber\\
&+\bigg(\beta  \mathcal{W} \epsilon+\frac{7}{12} \alpha  \eta' \mathcal{L}'+\frac{2}{3} \alpha  \mathcal{L} \eta''\nonumber\\
&+\frac{\eta^{(4)}}{24}+\alpha ^2 Y    \mathcal{L}^2+\frac{1}{6} \alpha \eta \mathcal{L}''\bigg)\Wt_{-2}\\
&+\left(-\frac{5}{3} \alpha  \mathcal{L} \eta'-\frac{\eta^{(3)}}{6}-\frac{2}{3} \alpha  \eta \mathcal{L}'\right)\Wt_{-1}\nonumber\\
&+\left(\frac{\eta''}{2}+2 \alpha  \eta \mathcal{L}\right)\Wt_0\Bigg]b.\nonumber
\end{align}
By inserting this gauge parameter into the field transformation equation given in (\ref{gauge transformation}), we derive the corresponding infinitesimal gauge transformations:
\begin{align}\label{ebc086}
\delta_{\lambda}\mathcal{L}&=\frac{k\epsilon^{(3)}}{4}+2\mathcal{L}\epsilon'+\epsilon\mathcal{L}'+2\eta\mathcal{W}'+3\mathcal{W}\eta'\\
\delta_{\lambda}\mathcal{W}&=\epsilon\mathcal{W}'+3\mathcal{W}\epsilon'+\frac{32}{15k}\Big(\mathcal{L}^2\eta'+\eta\mathcal{L}\mathcal{L}'\Big)+\frac{k\eta^{(5)}}{120}\nonumber\\
&+\frac{1}{15}\Big(3\eta'\mathcal{L}''+5\eta''\mathcal{L}'\Big)
+\frac{1}{10}\Big(\eta\mathcal{L}^{(3)}+5\eta^{(3)}\mathcal{L}  \Big)\\
\delta_{\lambda}\cX&=\frac{1}{10}\Big(\eta''\cY'-\eta'\cY''\Big)-\frac{1}{15}\Big(\eta^{(3)}\cY-\eta\cY^{(3)}\Big)\nonumber\\
&-\frac{32}{15k}\Big(\cY\mathcal{L}\eta'-\eta\mathcal{L}\cY'\Big)-\cX\epsilon'+\epsilon\cX'\\
\delta_{\lambda}\cY&=-\cX\eta'+2\eta\cX'-2\cY\epsilon'+\epsilon\cY'
\end{align}
For the prescribed boundary conditions, the boundary charges $\mathcal{Q}_{a}[\lambda]$ are well-defined and integrable, enabling us to associate $\mathcal{L}$ and $\mathcal{W}$ with the corresponding conserved charges. Similarly, the boundary charges $\mathcal{Q}_{x}[\lambda]$ are well-defined and integrable, allowing us to relate $\mathcal{X}$ and $\mathcal{Y}$ to the conserved charges.
Notably, these gauge transformations shed light on the structure of the asymptotic symmetry algebra \cite{Banados:1998pi}. 
By analyzing these asymptotic symmetries, one can determine the integral form of the variation of the canonical boundary charges, 

\begin{equation}\label{boundaryco11111333333}
    \mathcal{Q}_a[\lambda] = \int \mathrm{d}\varphi\; \left(\mathcal{L} \epsilon +\mathcal{W}\eta \right), \quad
    \mathcal{Q}_x[\lambda] = \int \mathrm{d}\varphi\; \left(\mathcal{X} \epsilon +\mathcal{Y}\eta\right).
\end{equation}
These canonical boundary charges serve as a fundamental tool in describing the asymptotic operator product expansion for the conformal boundary:
\begin{align}
  \mathcal{L}(\tau_1)\mathcal{L}(\tau_2) \sim& \frac{{3k}}{{2\tau_{12}^{4}}} + \frac{2\mathcal{L}}{{\tau_{12}^{2}}} + \frac{\mathcal{L}'}{\tau_{12}},\label{ope22}\\
   \mathcal{L}(\tau_1)\mathcal{W}(\tau_2) \sim&   \frac{2\mathcal{W}}{{\tau_{12}^{2}}} + \frac{\mathcal{W}'}{\tau_{12}},\label{ope23}\\
    \mathcal{W}(\tau_1)\mathcal{W}(\tau_2) \sim& \frac{{k}}{{\tau_{12}^{6}}} + \frac{2\mathcal{L}}{{\tau_{12}^{4}}} + \frac{\mathcal{L}'}{\tau_{12}^{3}}\label{ope33}\nonumber\\
 +&\frac{1}{\tau _{12}^2}\bigg(\frac{32\mathcal{L}^2}{15 k}+\frac{3\cL''}{10}\bigg)+\frac{1}{\tau _{12}}\bigg(\frac{32 \mathcal{L} \cL'}{15 k}+\frac{\cL^{(3)}}{15}\bigg),
 \end{align} 
These operator product expansions define the classical $\mathcal{W}_3$ algebra, which extends the Virasoro algebra by incorporating spin-3 operator $\mathcal{W}$. 
The operator product \eqref{ope23} illustrates how spin-2 and spin-3 modes interact, while the product \eqref{ope33} includes nonlinear contributions reflecting 
higher-spin interactions. As for the dilaton interactions:

\begin{align}  
\cL(\tau_1)\cX(\tau_2) \sim& - \frac{\cX}{{\tau_{12}^{2}}} + \frac{\cX'}{\tau_{12}},\\
\cL(\tau_1)\cY(\tau_2) \sim& - \frac{2\cY}{{\tau_{12}^{2}}} +\frac{\cY'}{\tau_{12}},\\
    \mathcal{W}(\tau_1)\cX(\tau_2) \sim& - \frac{2\cY}{{5\tau_{12}^{4}}} + \frac{\cY'}{5\tau_{12}^{3}}\nonumber\\
-&\frac{1}{\tau _{12}^2}\bigg(\frac{32\cY\mathcal{L}}{15 k}+\frac{\cY''}{10}\bigg)+\frac{1}{\tau _{12}}\bigg(\frac{32 \mathcal{L} \cY '}{15 k}+\frac{\cY^{(3)}}{15}\bigg),\\
\cW(\tau_1)\cY(\tau_2) \sim& - \frac{\cX}{{\tau_{12}^{2}}} + \frac{2\cX'}{\tau_{12}}.
  \end{align}
\vspace{0.5cm}

These structures provide a natural language for organizing asymptotic symmetries in $sl(3,\mathbb{R})$ dilaton gravity and play a key role 
in determining the boundary dynamics in the presence of symmetry breaking. While the above relations describe the undeformed $\mathcal{W}_3$ algebra, the inclusion 
of dilaton fields $\cX$ and $\cY$ in our extended $sl(3,\mathbb{R})$ model modifies this structure. These fields introduce additional terms in the algebra via boundary 
constraints, leading to symmetry breaking and a deformation of the standard commutation relations. The resulting structure retains the core features of $\mathcal{W}_3$ 
while encoding the dynamical effects of the dilaton sector.
\vspace{0.5cm}

From a physical perspective, the $\mathcal{W}_3$ algebra governs the dynamics of spin-2 and spin-3 boundary currents. The generator $\cL$ corresponds to the energy-momentum tensor, while $\cW$ represents a conserved higher--spin current of conformal weight three. In the holographic context, nontrivial $\mathcal{W}_3$--algebra charges reflect the presence of spin-3 hair at the AdS$_2$ boundary, leading to higher--derivative deformations in the dual theory. The inclusion of dilaton fields $\cX$ and $\cY$ modifies the standard $\mathcal{W}_3$ algebra by introducing symmetry-breaking terms, thus encoding how higher-spin symmetries are partially broken due to boundary conditions. This deformation mirrors the mechanism seen in the Schwarzian limit of $sl(2,\mathbb{R})$ JT gravity, now generalized to spin-3 dynamics.
\vspace{0.5cm}
\color{black}

It is clear that $\cX$ and $\cY$ behave as two boundary vectors, whereas $\cL$ and $\cW$ represent spin--2 and spin--3 densities. 
In this study, we have reformulated the boundary conditions in a way that eliminates any explicit dependence on $\lambda$. Consequently, the residual symmetries related to $a_t$ and $x$ correspond to circle diffeomorphisms generated by the vector fields $\epsilon$ and $\eta$ . Moreover, for the conformal boundary conditions in $sl(3,\mathbb{R})$ dilaton gravity, the resulting asymptotic symmetry algebra is given by a single copy of the $W_3$--algebra with a central charge of $c=3k$.
\begin{table}[ht]
    \centering
    \begin{tabular}{|c|c||c|c|}
        \hline
        charge  & \textbf{$\Delta$} &  dilaton  & \textbf{1$-\Delta$} \\
        \hline
         $\cL$ & 2 & $\cX$ & $-$1  \\
         \hline
         $\cW$ & 3 & $\cY$ & $-$2  \\  
         \hline
    \end{tabular}
    \caption{Conformal spins for conformal charges and related dilatons.}
    \label{tab:symbols}
\end{table}
\vspace{0.5cm}

 In the context of $sl(2,\mathbb{R})$ JT dilaton gravity, as well as in its $sl(3,\mathbb{R})$ extension with conformal boundary conditions, the asymptotic symmetry algebra at the boundary is initially identified as the infinite-dimensional $W_3$--algebra. However, the presence of the time-dependent dilaton field reshapes the asymptotic algebraic structure; as a result, even the expected preservation of the $SL(3,\mathbb{R})$ subgroup within the full $W_3$--symmetry is dynamically reduced, and the remaining infinite-dimensional symmetry components are broken. The dilaton acts as a stabilizer for the $sl(3,\mathbb{R})$ gauge connection and interacts with the gauge connection through boundary fluctuations, preventing the complete preservation of the infinite-dimensional $W_3$--symmetry. Table 1 presents the conformal charges along with the corresponding dilaton spin values. This symmetry reduction once again stems from the fluctuations of the dilaton field at the boundary. As a consequence, the residual symmetries related to $a_t$, $x$, and $y$ manifest as circle diffeomorphisms governed by the vector fields $\epsilon$ and $\eta$.
\vspace{0.5cm}

Thus, we identify the asymptotic symmetry algebra as the extended classical \( \mathcal{W}_3 \) algebra, rather than its quantum version. This classical structure arises naturally in our model and is consistent with the symmetry-breaking mechanism induced by the dilaton. So far, our discussion has been limited to this classical realization of the extended \( \mathcal{W}_3 \) symmetry. The quantum version, however, has already been discussed in equations \eqref{Q1}-\eqref{Q2}. One may naturally ask whether the quantum theory also admits a formulation in terms of a classical \( \mathcal{W}_N \) symmetry or its quantum analog. While these questions are certainly non-trivial, preliminary indications suggest that the procedure carries over in a consistent manner.

\section{Conclusions and Discussion}
\label{sec:conclusion}
\vspace{0.5cm}

Two dimensional gravity theories, particularly JT gravity where the dilaton field plays a dynamic role, are of significant interest in both theoretical physics and holography. \( sl(2,\mathbb{R}) \) JT gravity defines the asymptotic boundary conditions of AdS$_2$ geometry, enabling the spacetime diffeomorphism symmetries to extend into an infinite--dimensional Virasoro algebra. However, the presence of the time-dependent dilaton field reshapes the asymptotic algebraic structure; as a result, even the expected preservation of the $SL(2,\mathbb{R})$ subgroup within the full Virasoro symmetry is dynamically reduced, and the remaining infinite--dimensional symmetry components are broken. This breaking is associated with the fluctuations of the dilaton field at the boundary, which determine the effective dynamics of physical quantities at the boundary. In this context, the dilaton’s presence redefines the metric structure not only as a geometric but also as a dynamic parameter.
\vspace{0.5cm}

The \( sl(3,\mathbb{R}) \) JT gravity presents a more intricate framework as a natural extension of the classical \( sl(2,\mathbb{R}) \) theory. This theory incorporates higher--order gauge connections and a broader structure of asymptotic boundary symmetry algebras. The \( sl(3,\mathbb{R}) \) gauge connections introduce a new mechanism of symmetry breaking that does not even fully guarantee the preservation of the \( SL(3,\mathbb{R}) \) subgroup, significantly impacting the dynamics at the boundary. The \( sl(3,\mathbb{R}) \) structure allows for a wider perspective on holographic duality, where the effects of extended gauge symmetries on boundary quantities like energy, entropy, and other thermodynamic properties play a critical role in understanding both thermodynamic and holographic consequences of the theory.
Unlike earlier formulations of $\mathfrak{sl}(3,\mathbb{R})$ dilaton gravity based on coadjoint orbit techniques or metric formulations, our BF-theoretic framework systematically derives the full $\mathcal{W}_3$ structure and reveals novel dynamical effects induced by the dilaton fields, which have no counterpart in the $\mathfrak{sl}(2,\mathbb{R})$ setting.
\vspace{0.5cm}

The primary distinction between \( sl(2,\mathbb{R}) \) and \( sl(3,\mathbb{R}) \) JT gravities lies in the role of the dilaton field and its influence on the extension of asymptotic symmetry algebras. While the \( sl(2,\mathbb{R}) \) structure suffices to represent the isometric properties of AdS$_2$ geometry, the \( sl(3,\mathbb{R}) \) structure offers a more comprehensive mathematical framework to explore complex holographic dualities and the evolution of boundary physical quantities. Specifically, \( sl(3,\mathbb{R}) \) serves as a starting point for systematically extending higher--rank Lie groups and Chern--Simons--based approaches.
\vspace{0.5cm}

The findings presented in this work demonstrate that the asymptotic symmetry algebra is shaped not only by the structural properties of the gauge fields but also by the dynamical configurations of the dilaton field at the boundary. The dilaton does not constitute an external matter sector; rather, it is an intrinsic and structural component of the theory, taking values in the adjoint representation of the gauge algebra. Under appropriate boundary conditions, the time-dependent components of the dilaton become dynamical and contribute an abelian ideal, formed by mutually commuting modes, to the asymptotic symmetry algebra. This extension does not result from the inclusion of an external \( U(1) \) symmetry, but instead emerges from internal degrees of freedom that become visible at the boundary. Accordingly, the dilaton--induced commuting currents should be regarded as a natural substructure within the extended asymptotic symmetry algebra, supporting a semi-direct sum structure. The boundary behavior of the dilaton thus plays a decisive role in enriching the algebra while preserving its structural coherence, thereby revealing the significance of internal gauge dynamics in determining the asymptotic structure of two-dimensional gravitational systems.
\vspace{0.5cm}

In conclusion, the presence of the dilaton field significantly influences not only the physical dynamics at the boundary in both \( sl(2,\mathbb{R}) \) and \( sl(3,\mathbb{R}) \) JT gravities but also the extension and breaking of asymptotic symmetry algebras. These extended theories provide powerful tools for discovering new symmetry structures associated with boundary conditions, enriching the concept of holographic duality. In particular, these theories play a critical role in the detailed exploration of holographic frameworks such as the AdS/CFT correspondence and SYK model. Future research into higher--rank Lie group--based structures could further enhance the theoretical understanding in this field.
\paragraph{Comparison with Related Work:} As mentioned in the Introduction, several recent works have explored the interplay between generalized JT gravity and higher-spin extensions, particularly in connection with deformed SYK models and extended symmetry algebras. In ~\cite{Momeni:2024ygv} , a deformed SYK model including spin-3 operators is proposed as a potential holographic dual to generalized dilaton gravity. Our $\mathfrak{sl}(3,\mathbb{R})$ construction, with its explicit $W_3$ symmetry structure and controlled symmetry breaking, provides a concrete realization of this duality scenario. Furthermore, \cite{Momeni:2020tyt,Momeni:2020zkx,Momeni:2021jhk} develop complementary approaches based on coadjoint orbit methods, quantization of extended $W$-algebras, and generalized boundary dynamics. While sharing similar motivations, our work distinguishes itself through a BF-theoretic formulation that systematically derives the asymptotic symmetry algebra and clarifies the role of dilaton-induced symmetry breaking. This perspective contributes to a deeper understanding of how higher-spin structures can be implemented in two-dimensional holography.
\vspace{0.5cm}

In our model, the interactions of the dilaton and higher-spin fields at the boundary lead to the breaking of asymptotic symmetries. The initially present infinite-dimensional $\mathcal{W}_3$ algebra is reduced to its $\mathrm{SL}(3,\mathbb{R})$ subalgebra due to the effect of the dilaton. This process bears similarity to the Schwarzian-style symmetry breaking. However, in addition to this complete symmetry breaking, the fluctuations of the dilaton at the boundary also allow for the \textbf{extension} of the symmetry in some cases. As a result, our model exhibits both \textbf{asymptotic symmetry breaking} and \textbf{symmetry extension} occurring simultaneously. See Appendix~\ref{sec:appB} for a summary of boundary theories and their symmetry structures.
\vspace{0.5cm}

At the end, JT gravity as a two-dimensional BF theory provides complementary frameworks for understanding the geometric and physical origins of asymptotic symmetries and Schwarzian dynamics. The gauge symmetries of BF theory, together with the algebraic structure of the target space, underpin the asymptotic symmetry algebras at the boundary, while JT gravity serves as a specific realization of these structures. In both theories, the Schwarzian action emerges as an effective reduced description of residual symmetries arising from bulk gauge transformations. However, the Schwarzian dynamics do not encompass the full gauge freedoms but instead capture the dynamics of a physically meaningful subset. Thus, BF theory and JT gravity offer a rich and complementary approach to elucidating the complete structure of boundary symmetries and their holographic implications.
\vspace{0.5cm}

It remains an open question whether the the $sl(3,\mathbb{R})$ JT gravity framework presented here admits a well--defined quantum mechanical dual--perhaps in the spirit of the SYK--model, possibly incorporating spin$-3$ operators or deformed reparametrization modes--and whether it allows for a consistent supersymmetric extension along the lines of $\mathfrak{osp}(2|\cN)$  BF theories, where super--Schwarzian dynamics are expected to emerge at the boundary  \cite{Cardenas:2018krd}.

\newpage

\section{Acknowledgments}
\label{sec:ack}
\textit{
Authors are supported by The Scientific and Technological Research Council of Türkiye (TÜBİTAK) through the ARDEB 
1001 project with Grant number 123F255. 
}
\appendix
\section{Notation Table}
\label{sec:appa}

\begin{table}[h!]
\centering
\begin{tabular}{|c|l|}
\hline
\textbf{Symbol} & \textbf{Description} \\
\hline
$\Lt_i, \Wt_i$ & Elements of the $\mathfrak{sl}(2,\mathbb{R})$ and $\mathfrak{sl}(3,\mathbb{R})$ algebras \\
\hline
$\mathcal{L}^i,\mathcal{W}^i$, $\mathcal{X}^i, \mathcal{Y}^i$ & Affine boundary charges and dilaton components \\
\hline
$\epsilon^i, \eta^i$ & Affine boundary variation parameters \\
\hline
$\mathcal{L},\mathcal{W}$, $\mathcal{X}, \mathcal{Y}$ & Conformal boundary charges and dilaton components \\
\hline
$\epsilon, \eta$ & Conformal boundary variation parameters \\
\hline
\end{tabular}
\caption{Notation Table}
\end{table}

\section{Summary of Boundary Theories}
\label{sec:appB}

\begin{table}[h!]
\centering
\begin{tabular}{|c|l|c|}
\hline
\textbf{Dilaton gravity theory} & \textbf{Boundary fields} & \textbf{Asymptotic symmetry algebra} \\
\hline
\(\mathfrak{sl}(2,\mathbb{R})\)-BF (affine) & $\mathcal{L}^i, \mathcal{X}^i$ & Extended affine $\mathfrak{sl}(2,\mathbb{R})_k$ \\
\hline
\(\mathfrak{sl}(2,\mathbb{R})\)-BF (conformal) & $\mathcal{L}, \mathcal{X}$ & Extended $\mathcal{W}_2$ \\
\hline
\(\mathfrak{sl}(3,\mathbb{R})\)-BF (affine) & $\mathcal{L}^i, \mathcal{W}^i, \mathcal{X}^i, \mathcal{Y}^i$ & Extended affine $\mathfrak{sl}(3,\mathbb{R})_k$ \\
\hline
\(\mathfrak{sl}(3,\mathbb{R})\)-BF (conformal) & $\mathcal{L}, \mathcal{W}, \mathcal{X}, \mathcal{Y}$ & Extended $\mathcal{W}_3$ \\
\hline
\end{tabular}
\label{tab:boundary-summary}
\caption{
\shortstack{
Boundary dynamics and asymptotic symmetries for \(\mathfrak{sl}(2,\mathbb{R})\) and\\
 \(\mathfrak{sl}(3,\mathbb{R})\) dilaton gravity with affine and conformal boundary conditions.
}
}
\end{table}


\end{document}